\documentclass[12pt,preprint]{aastex}
\usepackage{graphicx}

\begin{document}
\title{Rapid cooling of neutron star in Cassiopeia A and $r$-mode damping in the core}

\author{Shu-Hua Yang\altaffilmark{1}
\email{ysh@phy.ccnu.edu.cn}}
\author{Chun-Mei Pi\altaffilmark{2}}
\author{Xiao-Ping Zheng\altaffilmark{1}
\email{zhxp@phy.ccnu.edu.cn}} \altaffiltext{1}{Institute of
Astrophysics, Huazhong Normal University, Wuhan, 430079, China}
\altaffiltext{2}{Department of Physics and Electronics, Hubei University of Education, Wuhan 430205, China}

\begin{abstract}
We proposed alternative explanation to the rapid cooling of neutron
star in Cas A. It is suggested that the star is experiencing the
recovery period following the \textit{r}-mode heating process,
assuming the star is differentially rotating. Like the
neutron-superfluidity-triggering model, our model predicts the rapid
cooling will continue for several decades. However, the behavior of
the two models has slight differences, and they might be
distinguished by observations in the near future.
\end{abstract}


\section{Introduction}

The neutron star in Cassiopeia A (Cas A) is one of the most
important isolated neutron stars (NSs) in testing the thermal
evolution theory of NS because both its age and surface temperature
are reliably estimated: $t \approx 330\pm 20$~yr (Fesen et al. 2006)
and $T_{s} \sim 2 \times 10^6$~K (Ho \& Heinke 2009). The importance
of Cas A NS was greatly enhanced recently, as Ho \& Heinke (2010)
found a steady decline of $T_{s}$ by about 4\% by analyzing the 10
year $\textit{Chandra}$ observations of it, and the new
observational data reported by Shternin et al. (2011) confirms and
extends this cooling trend (see table 1 in Shternin et al. 2011).

Page et al. (2011) and Sheternin et al. (2011) suggested that the
observed decreasing of the surface temperature of Cas A NS is
difficult to explain by the cooling theory in spite of considering
the triplet-state neutron superfluidity: this cooling rate is much
larger than expected from the standard modified Urca process; and it
is not likely due to the crust-core relaxation, which is supposed to
last for typically $\leq100$ yr (e.g., Lattimer et al. 1994;
Yakovlev et al. 2010). They also found that if this rapid cooling is
triggered by "breaking and formation of Cooper pairs (PBF)"
neutrino-emission process, the cooling data may constrain the
critical temperature of the triplet-state neutron superfluidity to
several times of $10^8$K.

In this \emph{Letter}, we present an alternative explanation to the
rapid cooling of Cas A NS, which suggests that the star is
experiencing the recovery period when the \textit{r}-mode heating
process if over. In the next section, we discuss the \textit{r}-mode
heating mechanism during the thermal evolution of neutron stars.
After that, our explanation of the cooling data of Cas A NS is
given. The last section is our conclusions and discussions.

\section{\textit{R}-mode heating in neutron stars}
Ever since 1998, \textit{r}-mode instability is extensively studied
in compact stars as the most important gravitational
radiation-driven Chandrasekhar-Fridman-Schutz instability (Andersson
1998; Friedman \& Morsink 1998), it is believed that it determines
the spin limit of compact stars and the gravitational wave emitted
during the instable process of the star can be detected by the new
generation of gravitational-wave detectors (Andersson \& Kokkotas
2001). However, The role of the \textit{r}-mode dissipation for the
thermal evolution of compact stars has long been ignored because the
heating effect due to the \textit{r}-mode dissipation is supposed to
exist only in the first several decades of the newly born neutron
stars (Watts \& Andersson 2002). Nevertheless, Zheng et al. (2006)
found that for strange stars made of strange quark matter (SQM), the
\textit{r}-mode heating effect can last for even $10^7$ years.

In fact, there exists a saturated amplitude for \textit{r}-modes in
compact stars during the instable process. Since the saturated
amplitude is determined by the nonlinear effects, it is usually put
into the model artificially (Owen et al. 1998; Ho \& Lai 2000). Many
efforts have been paid to the study of the nonlinear effects to give
a saturated \textit{r}-mode amplitude naturally (e.g., Schenk et al.
2002; Arras et al. 2003; Brink et al. 2004, 2005). As an important
nonlinear effect, differential rotation induced by \textit{r}-modes
was studied extensively (Rezzolla et al. 2000, 2001a,b; Stergioulas
\& Font 2001; Lindblom et al. 2001, S\'a 2004).  Among which S\'a
(2004) and S\'a \& Tom\'e (2005, 2006) solved the fluid equations
within nonlinear theory up to the second order in the mode amplitude
and described the differential rotation analytically. By doing so,
they obtained a saturated amplitude of \textit{r}-modes
self-consistently, which depends upon the parameter that describing
the initial condition of the differential rotation.

Using the model developed by S\'a (2004) and S\'a \& Tom\'e (2005,
2006), Yu et al. (2009) investigated the long-term spin and thermal
evolution of isolated NSs under the influence of the differential
rotation, and pointed out that the stars can keep nearly a constant
temperature for over a thousand years since the differential
rotation significantly prolongs the duration of \textit{r}-modes.
The detailed study by Yang et al. (2010) found that the heating
effect of the prolonged \textit{r}-modes enables us to explain the
two young and hot pulsar's (PSR B0531+21 and RX J0822-4300)
temperature data with NS model composed of simple $npe$ matter,
without the inclusion of superfluidity or exotic particles.

Both in the thermal evolution curves of fig.2 in Yu et al. (2009)
and fig.1 in Yang et al. (2010), It can be easily found that a rapid
cooling period emerges immediately after the \textit{r}-mode heating
process is switched off. Therefore, we try to explain the observed
rapid cooling of Cas A NS following Yang et al. (2010) in the next
section.

\section{The results}

In order to simulate the thermal evolution, the thermal evolution
equation of NS must be solved numerically coupling with the
equations of the \textit{r}-mode evolution and the spin evolution of
the star(see equations (13), (14) and (15) of Yang et al. (2010)).
We take the initial temperature $T_0=10^{10}$K, the initial
\textit{r}-mode amplitude $\alpha_0=10^{-6}$ and the initial angular
velocity $\Omega_0=\frac{2}{3}\sqrt{\pi G \bar{\rho}}$.

In our simulation, we took the magnetic field as $B=5\times10^{10}$
G since the X-ray spectral fits of Cas A NS suggest $B<10^{11}$G (Ho
\& Heinke 2009), and this NS is believed to be one of the several
so-called central compact objects (CCOs) which have  $B \sim
10^{10}-10^{11}$G (Halpern \& Gotthelf 2010; Ho 2011).

Following Yang et al. (2010), a moderately stiff equation of state
(EOS) proposed by Prakash et al. (1988) is employed (model I). The
maximum mass of this model is $M=1.977M_{\odot}$, and the direct
Urca process is forbidden in the case $M<M_{D}=1.36M_{\odot}$. The
relation between $T_{s}$ and the internal NS temperature $T$ is
taken from Potekhin et al. (1997), which supposed the outer heat
blanketing NS envelope is made of ion and neglected the effects of
surface magnetic fields.

The cooling data of Cas A NS is taking from table 1 of Shternin et
al. (2011). Mention that the effective surface temperature detected
by a distant observer is $T_{s}^{\infty}=T_{s}\sqrt{1-R_{g}/R}$,
where $R_{g}$ is the gravitational stellar radius. Since we mainly
focus on the $M=1.361M_{\odot}$ neutron star model (the
corresponding radius is $12.93$ km), $T_{s}^{\infty}=0.83T_{s}$ is
taken.

Fig.1 shows the cooling curves of neutron stars with different
masses and fixed $K$=2($K$ is a free parameter describing the
initial condition of the differential rotation). The plateaus of the
curves indicate the heating effect due to \textit{r}-mode
dissipation, and the duration of this high temperature depends on
the parameter $K$ for selected neutron star mass (see Fig.3 of Yang
et al. 2010). Since the direct Urca process begins to happen in the
$M=1.36M_{\odot}$ neutron star model (which can greatly enhance the
neutrino emission rate comparing with the modified Urca process),
the cooling curves depend on the mass sensitively in the vicinity of
it (see the curves of $M=1.361M_{\odot},
1.362M_{\odot},1.365M_{\odot}$), and only the $M=1.361M_{\odot}$
curve passes through the region where the observed data located.
Comparing to Fig.2 of Yu et al. (2009), one can easily find that the
rapid cooling which can be used to explain the Cas A NS data occurs
just after the completely shutoff of the \textit{r}-mode heating
process.

Fig.2 displays the cooling curves of the $1.361M_{\odot}$ neutron
star with different values of $K$ and the curves are compared with
the Cas A NS data. In comparison with the observations, the
differential rotation parameter $K$ can take the values around 2.0.
In Fig.3 we plot our best fitted cooling curve ($K$=2.3) in a larger
span of ages and the curve without the \textit{r}-mode heating
effect is also displayed for comparison. An insert is also displayed
to show the possible temperature drop in the following twenty years
and the grey rectangle indicates the possible temperature scope
predicted by the neutron-superfluidity-triggering model. It can be
seen that the rapid cooling near the Cas A NS data will continue for
several decades, and it will take a few hundred years to recover to
the cooling rate of that not considering the \textit{r}-mode heating
effect. Although the behavior of the rapid cooling process of our
model seems similar to neutron-superfluidity-triggering model (Page
et al. 2011; Sheternin et al. 2011), they are different in detail.
The part of the curve which best fits the observational data in both
of their studies (see Fig.3 of Page et al. (2011) and Sheternin et
al. (2011)) are closer to straight line than that of us. As a
result, our best fitted curve predicts about 2\% higher temperature
in two decades (see the insert). Perhaps, this difference is
distinguishable in future observations.

In fig.4, we plot the evolution of the amplitude of r-modes with the
same parameters as fig.3. Abadie et al. (2010) analyzed the data of
Cas A NS in a 12 day interval taken by LIGO, no gravitational-wave
signal is found. But they gave the upper limits on the r-mode
amplitude, which is 0.005-0.14. As far as our model is concerned,
one can easily seen from fig.4 that the amplitude of r-modes has
dropped to 0.005 about 24 years ago (that is the Year 1987). We also
don't expect the gravitational-wave due to r-modes can be observed
even by the advanced LIGO and Virgo interferometers, since the
amplitude of r-modes has declined to its initial
value($\alpha=10^{-6}$) in 1997.

\section{Conclusions and discussions}

Based on our former work about the \textit{r}-mode heating effect of
NS, we explained the rapid cooling of Cas A NS. When the
\textit{r}-mode heating process is switched off, the NS cools down
rapidly, and it needs a few hundred years to recover to the normal
cooling rate (which refers to the case that don't consider the
\textit{r}-mode heating effect). The rapid cooling as the Cas A NS
occurs in the beginning of this recovery period and it will last for
several decades. We also found that the behavior of our model and
the neutron-superfluidity-triggering model has slight differences,
and they might be distinguished by observations in the near future.

The essential element of our interpretation to the rapid cooling of
Cas A NS is that the \textit{r}-mode instable period of the star can
last for about 300 years. Although many models about the nonlinear
evolution of \textit{r}-modes in NSs didn't expect such a long
duration of \textit{r}-mode instability(e.g., Rezzolla et al.
2001a,b; Bondaresu et al. 2009), some models did support it(e.g.,
Arras et al. 2003).

In our calculation, we assumed the outer heat blanketing NS envelope
is made of ion. However, the Cas A NS is believed to have a carbon
atmosphere (Ho \& Heinke 2009). Theoretically, light elements make
the envelope more heat transparent and the surface temperature of
the same mass NS can be risen (Potekhin et al. 1997), and one has to
explain the observations of the Cas A NS with the NS model of a
little larger mass than $1.361M_{\odot}$.

What's more, we only considered a constant magnetic field in our
simulation. Nevertheless, Rezzolla et al. (2001a,b) showed that the
differential rotation induced by \textit{r}-mode will generate a
strong toroidal magnetic field, this could effect the
\textit{r}-mode evolution and should be taken into account in
further work.

\acknowledgements We would like to acknowledge Y.W. Yu for useful
discussion. We are especially indebted to the anonymous referee for
his/her useful comments that helped us to improve the paper. This
work is supported by the National Natural Science Foundation of
China (grant no. 11073008).

\begin{figure}
\plotone{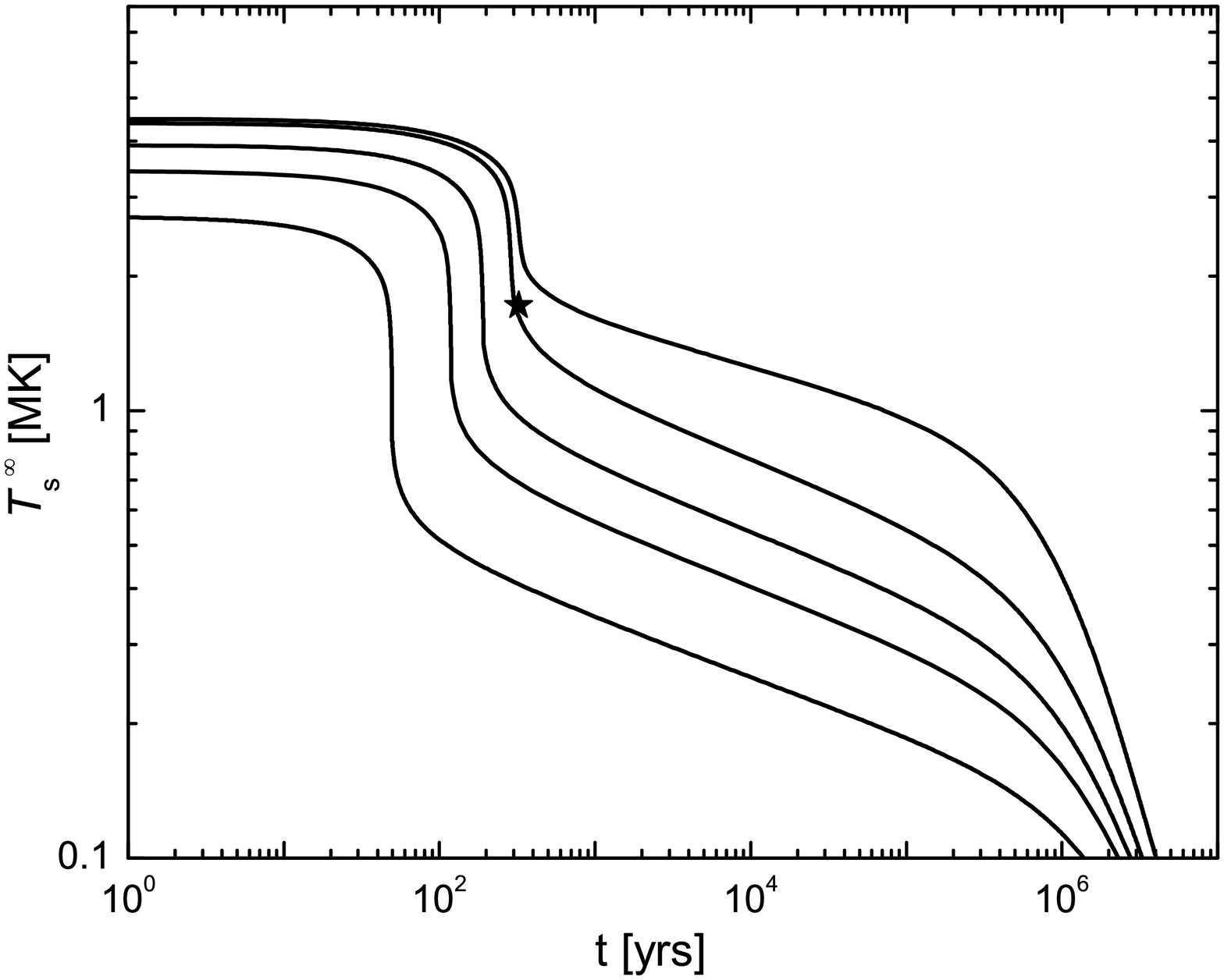}\caption{Cooling curves of neutron stars with
$K=2$. The curves correspond to the NS
   mass $1.360M_{\odot}$, $1.361M_{\odot}$, $1.362M_{\odot}$, $1.365M_{\odot}$ and $1.4M_{\odot}$,
   respectively. The pentagram presents the location of the observed cooling data of Cas A NS.} \label{figure:1}
\end{figure}

\begin{figure}
\plotone{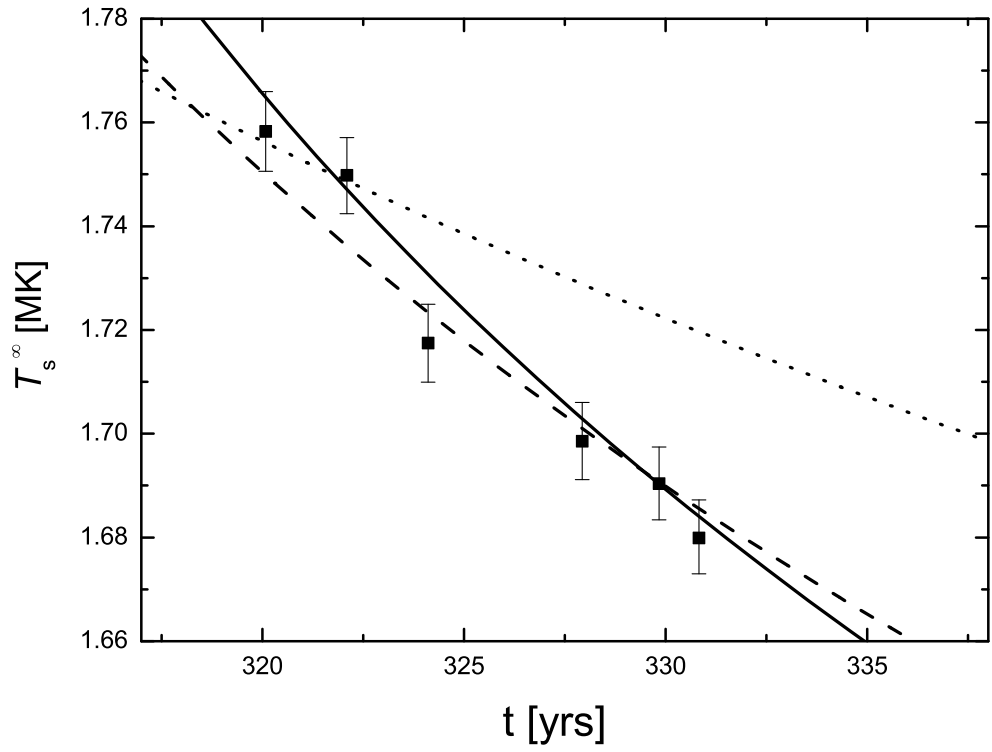}\caption{Cooling curves of the $1.361M_{\odot}$
neutron star. The dot, dashed and solid curves correspond to
$K=1.5$, $K=2.1$ and $K=2.3$, respectively.} \label{figure:2}
\end{figure}

\begin{figure}
\plotone{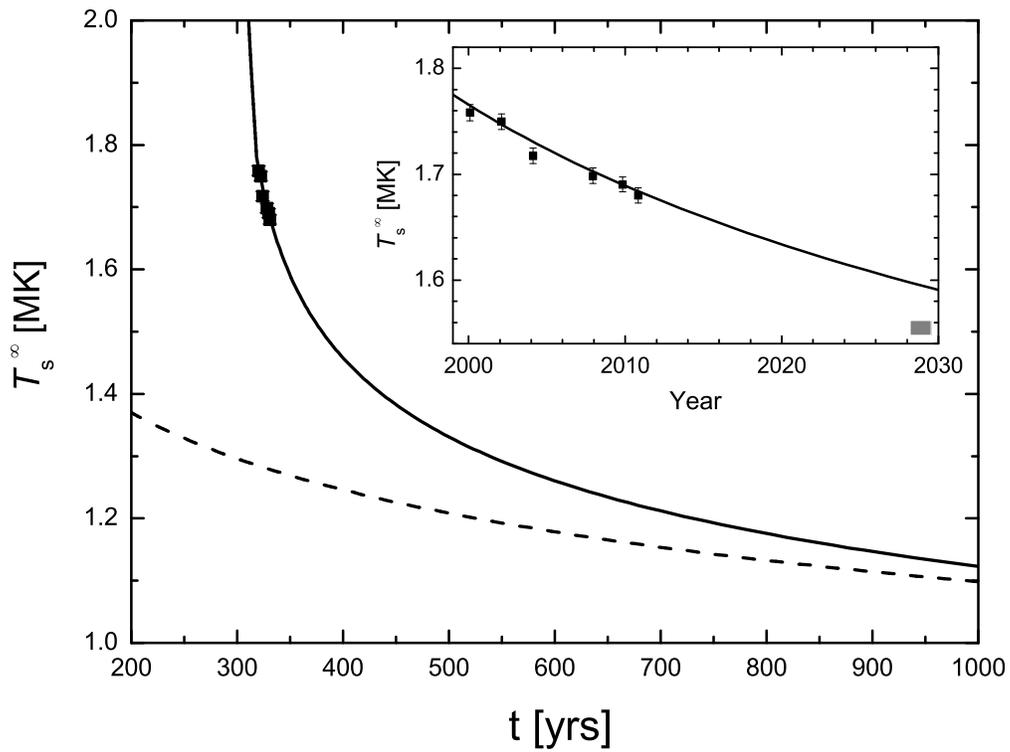}\caption{Cooling curves of the $1.361M_{\odot}$
neutron star with $K=2.3$ (solid line). For comparison, the dashed
line is calculated without the \textit{r}-mode heating effect. The
insert shows the temperature evolution in the following twenty years
and the grey rectangle indicates the possible temperature scope
predicted by the neutron-superfluidity-triggering model.}
\label{figure:3}
\end{figure}

\begin{figure}
\plotone{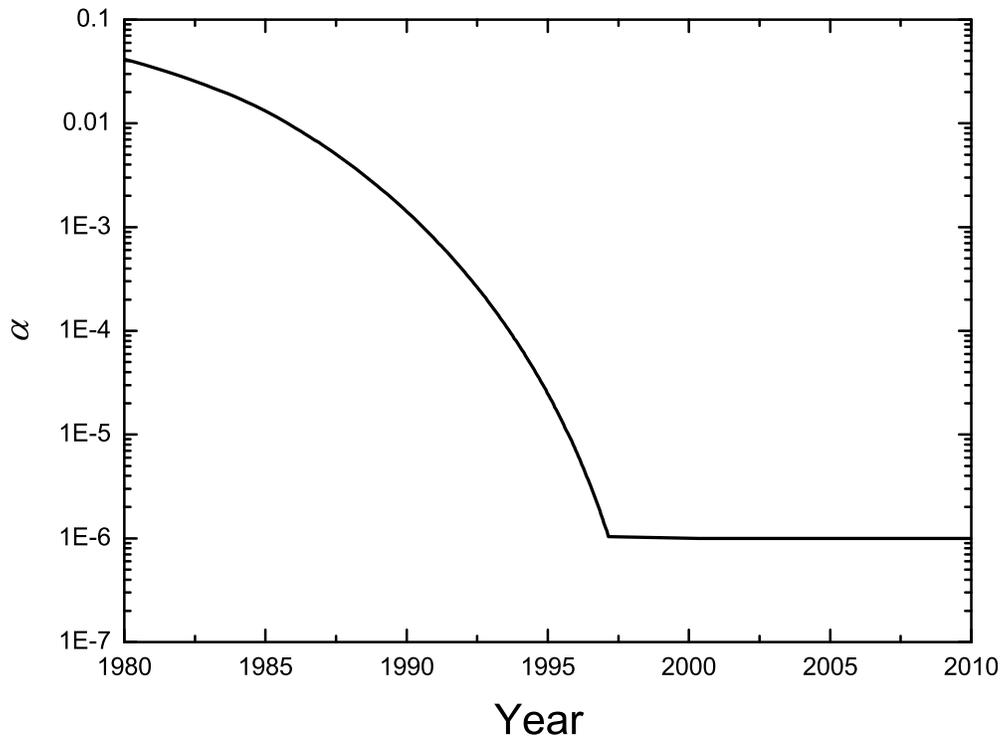}\caption{The evolution of the amplitude of r-modes
of the $1.361M_{\odot}$ neutron star with $K=2.3$.} \label{figure:3}
\end{figure}

\end{document}